# First Direct Observation of Nanometer size Hydride Precipitations on Superconducting Niobium

Z-H Sung[1], A Cano[1], A Murthy[1], E Karapetrova[2], J-Y Lee[1], M Martinello[1,3], A Grassellino[1] and A Romanenko[1]

1 Fermi National Accelerator Laboratory, Batavia, IL, 60510
2 Advanced Photon Source, Argonne National Laboratory, Lemont, IL, 60439
3 SLAC National Accelerator Laboratory, Menlo Park, CA, 94025

**Abstract:** Superconducting niobium serves as a key enabling material for superconducting radio frequency (SRF) technology as well as quantum computing devices. At room temperature, hydrogen commonly occupies tetragonal sites in the Nb lattice as metal (M)-gas (H) phase. When the temperature is decreased, however, solid solution of Nb-H starts to be precipitated. In this study, we show the first identified topographical features associated with nanometer-size hydride phase ($Nb_{1-x}H_x$) precipitates on metallic superconducting niobium using cryogenic-atomic force microscopy (AFM). Further, high energy grazing incidence X-ray diffraction reveals information regarding the structure and stoichiometry that these precipitates exhibit. Finally, through time-of-flight secondary ion mass spectroscopy (ToF-SIMS), we are able to locate atomic hydrogen sources near the top surface. This systematic study further explains localized degradation of RF superconductivity by the proximity effect due to hydrogen clusters.

Niobium (Nb) is a 4d transition bcc (body centered cubic) metal that acts as a marginal type II superconductor below $T_c$ (the critical temperature) of 9.2 K with a narrow-mixed state of magnetic vortices pinning[1,2]. It has the highest lower critical field, $H_{c1}(0) \approx 180$ mT, and the longest coherence length, $\xi_{Nb}(0) = 40$ nm, among superconductors [3,4]. Additionally, its extended low yield strength mechanical property allows to easily form into various desired geometries[5]. These unique properties make niobium an enabling materials applicable for 3D superconducting radio frequency (SRF) resonator (cavity)[5], superconducting quantum interference devices (SQUID)[6,7], and most recently for quantum computing devices[8,9].

In 3D SRF Nb cavity, various engineering treatments are employed to achieve high quality factor ($Q_0$), close to the maximum theoretically expected accelerating field, 50 MV/m[5]. For instance, baking the cavity at 120°C/48hrs[10,11] and doping the cavity by exposing it to nitrogen[12–14] are known to significantly reduce the surface resistance in the superconducting state and improve the overall cavity performance, associated with dirty-limit superconductivity on the nano-structured SRF surface[15,16]. However, hydrogen is highly soluble in the niobium upon cryogenic cooling[17]. Nanometer-size of Nb hydride ($Nb_{1-x}H_x$) can preferentially precipitate on the cavity surface[18,19]. As the hydride precipitate is a normal conductor at SRF cavity operation temperature, T ≈ 2 K, the presence of these precipitates lead to local breakdown of surface superconductivity and manifests high field Q slopes[10,20].

As these hydride precipitates only form at cryogenic temperature, direct *in-situ* observation is challenging. Analytical electron microscopy[21–23] presented local segregation of the nano-sized Nb-H phases on various SRF Nb cavities, but as this technique primarily provides local information from selected areas, alternative modalities are needed to obtain an understanding of the three dimensional morphology and distribution of these precipitates. The initial *in-situ* study[24,25] using laser confocal microscopy directly observed the formation of Nb hydride precipitations that left behind permanent physical surface deformation shown as μm-size seashell scars. Especially, the study found that nanometer scale hydride-like surface features appeared on the surface and completely vanished when heated back to room temperature. Its size is comparable to the coherence length of niobium, $\xi_{Nb}(0) \approx 40$ nm. However, this finding was limited only to the Nb sample that has been exposed to hydrogen rich environment.

In the SRF Nb, hydrogen concentrations are known to be highest in ~ 100 nm region closest to the surface[26,27], underneath the topmost $Nb_2O_5$ layer[28]. Understanding of the size, morphology, and distribution of the nanometer scale Nb hydride-like features requires the use of additional techniques at cryogenic temperature, which will be helpful for evaluating the extent of localized degradation in the cavity performance. Also, importantly, the structural phase of the surface precipitates is unexplored. Thus, in this study, by directly analyzing cut-out coupons from tested Nb SRF cavities, we further explore the morphology of the nano-size Nb hydride precipitates while identifying their structural phase through combination of cryogenic atomic force microscopy (AFM) and high energy grazing incidence x-ray diffraction (GI-XRD) technique. This systematic study provides insight into the role of proximity coupled Nb hydride precipitates that play in suppressing the overall RF superconductivity.

**Result**

**Surface topographical features on the SRF Nb upon cryogenic cooling**

The overview of a single cell SRF Nb cavity and the cryogenic-AFM system used in this study are represented in Fig. 1. The machined parts of the outside cavity depict local areas where thermal fluctuations occurred during RF test[29]. A 10 mm diameter cut-out coupon is placed on the top copper plate of the X-Y-Z piezo-stage in the AFM. The sample was cooled at a rate of 20 K/min, and contact-mode AFM scans were performed at every 50 K step from 300 K to 10 K and as the sample was warmed back up to 300 K. Due to thermal fluctuation of an AFM cantilever piezo, an individual scan was executed after allowing 3 hours for the system to stabilize at each target temperature. This stabilization period provides an opportunity for hydrogen atoms incorporated within the niobium lattice to form Nb-H phase with extended diffusion[30].

Table 1 describes the characteristics of the cavities and its cut-out coupons, investigated in this study. $Q_0$ (quality factor) vs $E_{acc}$, (accelerating voltage) of the cavities are represented in the supplementary Figure 1. Hydrogen-related performance degradations are seen in the cavities. The green arrow indicates a marked Q drop due to hydrogen-disease[31], and the black arrow points out high field Q slope (HFQS) associated with nano-sized Nb hydride precipitates[18]. Anti-Q slope behavior is seen on the N (nitrogen)-doped cavity[12]. 120°C mild baking resolves HFQS that appears on the 800°C HT'ed + EP cavity[10]. The N -infused cavity exhibits the highest performance close to 45 MV/m[14].

We first scanned a "hot spot" area of a non-degassed Nb cavity with cryo-AFM. This cavity was only extra 10 μm EP'ed after a standard ~40 μm EP reset process, meaning that there is no thermal effect on the cavity surface. Its RF performance presents earlier HFQS behavior, compared with an 800°C EP'ed cavity as shown in the supplementary Figure 2. The hot spot represents a local area where thermal heating occurred on the cavity surface during RF measurements at cryogenic temperature, which is indicative of significant thermal breakdown of superconductivity[32]. AFM laser intensity images at various temperatures are shown in Fig. 2. At 300 K and 250 K, no distinguishable features are seen beside scratches and/or foreign particles arising from sample handling. However, distinct surface features such as small bumps appeared during cooling to 150 K, as indicated by the red arrows. These surface features remained down to 10 K., which are ~ 800 nm in diameter and ~ 30-50 nm in height, as described in Fig. 2b and 2c. As the sample is warmed to room temperature, however, the surface features disappear, as shown in Figure 2d. Based on the similar phenomenon observed in the previous *in-situ* study[24,25], we suspect that these surface features arise from the formation of Nb hydride segregations. In this case, residual hydrogen in the hot spot region could be sufficient to drive Nb hydride precipitates on the nanometric scale, but insufficient to form larger precipitates on the order of a few hundreds of micrometers (μm) size that introduce surface scars.

Figure 3 compares the surface features observed between 300 K and 100 K in the cut-out coupons from the representatives of the SRF cavities (described as Table 1 and in the supplementary Figure 1). From the hot spot, several µm size features are observable at 200 K, and, as described before, nm-scale features started appearing at 150 K between the larger features. The 800°C/2hrs heat treatment is empirically known to annihilate interstitial hydrogens in the Nb surface[33]. We found that the post 800°C heat treated sample shows a markedly suppressed formation of the surface features. But many smaller sized features still appeared at 100 K. The 120°C baked and N (nitrogen)-doped cut out exhibits surface features even at 200 K. The latter, however, shows a large density of the features at 100 K. Interestingly, residual particles on the 120°C baked cut-out, marked with the green arrows, appears to increase in size from 300 K to 200 K upon cooling. It is assumed that structural defects localized near the surface may facilitate the formation of Nb hydride precipitates in this case. In contrast, the N (nitrogen)-infused cut-out shows no distinct surface features down to 100 K. In most cases, the surface protrusions that appeared up to at 100 K remained stable while cooling down to 10 K. Representative 7 µm by 7 µm area scans and 3 D topographical plots are provided in Figure 4. However, the 120°C baked and N-doped cut-outs have high density of the small features whose size are too small to ascertain within our AFM resolution limit.

The averaged frequencies (density) and heights of clearly observed surface features for each cut-out coupon are compared in Fig. 5. The averaged total number of the observed features is presented in the last column in Fig. 5a. The hot spot and post 800°C HT'ed cut-outs shows high densities of the protrusions, compared to the others, as shown in Fig. 5a. The hot spot exhibits large size features (> 100 nm) at 200 K, but on the 800°C HT'ed cut-out, the features with heights less than 10 nm are predominant. On the 120°C baked, N-doped, and N-infused cut-outs, these surface features are observed at a low density, and their heights are mainly between 10 nm $\ll$ h $\ll$ 50 nm. However, the surface features appeared at 250 K for the 120°C baked cut-out, which is higher temperature compared to that observed in other samples. The N-infused coupon shows the lowest density of the surface features.

**The phase of Nb-H precipitates and structural variation on the SRF Nb surface upon cryogenic cooling**

To verify that the surface features observed by cryo-AFM scan arise from hydride precipitations, the surfaces of the cut-out coupons were probed by high energy X-ray diffraction (XRD) at grazing incidence as a function of temperature. First, we investigated a heavy hydrogen-loaded (H-enriched) sample based on findings from the previous *in-situ* study[24,25]. In that study, distinct surface features associated with both of micro- and nano-scale Nb-H phase segregations were observed at temperature < 150 K, following to the Nb-H phase diagram[34]. The H-enriched sample was prepared in a same fashion to the initial study, but at this moment, a cut-out from a post 975°C/3hr HT'ed N-doped cavity that showed meaningful magnetic flux expulsion[35] was applied, as described in Method.

Figure 6 shows a series of 2θ patterns of high energy XRD at 1° of grazing incident angle as a function of temperature. The 2θ patterns reveal a general systematic edge reduction of the unit cell, $a$(300 K) = 0.3262 nm upon cryogenic cooling to 30 K, associated with a roughly 0.1% contraction of the Nb-matrix contraction. The collected patterns from 300 K to 200 K at every 25 K steps show variations in the intensity, associated with in the bcc Nb reflections, leading to a preferential orientation toward the (2 1 1) plane as well as an increase of the background at low angles (θ < 20°) close to the Nb (101) reflection. This could be due to the displacement of H atoms during the initial stages of Nb-H precipitate formation leading to Nb lattice distortion. At 175 K, new prominent peaks appear on the low angular site of (1 1 0) bcc-Nb peak, corresponding to secondary phase formations as shown in Fig. 6 inset (b). The peaks are identified as arising from the presence of cubic (bcc)-$NbH_{0.8}$ phase and orthorhombic (orth)-$NbH_{0.89}$. The bcc-Nb-H phase has a lattice parameter of a ≈ 0.342 nm and the orthorhombic Nb-H has lattice parameters of a = 0.484 nm, b = 0.49 nm and c= 0.345 nm, respectively. The schematic unit cell structure of bcc-$NbH_{0.8}$ and orth-$NbH_{0.89}$ phases are represented in Fig. 6 inset (a). At 175 K, the intensity ratio of the bcc-Nb-H : orth-Nb-H peak is 2.5 : 1, but it decreases to 1.2: 1 at 150 K. When the temperature is further decreased to 30 K, the angular positions of reflections corresponding to Nb-hydrides remain stable, but their intensities remarkably decrease and resemble subtle "shoulder" peaks, just above the background. The small peak located between the Nb-H phase and Nb (101) peak at 100 K and 30 K is not identified in this study. Similar to the observations by cryo-AFM, the reflections corresponding to Nb-hydride phases completely disappeared when warmed back to 300 K, and the reflections associated with bcc-Nb remain. After this GI-XRD study at cryogenic temperature, the H-enriched sample exhibits seashell-shape scars on the surface (in the supplementary Figure 3), which is analogous to the initial study[24,25].

After scans on the H-enriched sample, we investigated two additional cut-out coupons taken from cavities that exhibited RF performance degradation related to the likely presence of Nb hydride precipitates. These are a non-degassed and EP + 800°C/2hrs cavities showing high field Q-slope[10] (supplementary Figure 2). Figure 7 shows a series of the 2θ X-ray reflections recorded at 1° of grazing incidence from 300 K to 50 K for (a) the hot spot from the non-degassed and (b) the cut out of the EP + 800°C HT'ed cavity, respectively. Likewise, the primary reflections corresponding to the Nb matrix (bcc-Nb) showed a systematic shift towards a higher 2θ angle, indicating unit cell contraction upon cryogenic cooling, similar to Fig. 6. Starting at 200 K, the hot spot sample shows weak reflections corresponding to the bcc-$NbH_{0.8}$ phase at 2θ = 18.46 and 32.7 with d (interplanar distance) = 2.42 Å and 1.37 Å, respectively, and these peaks remain stable down to 50 K. The cut out of the EP + 800°C HT'ed cavity shows a different behavior, as compared with the insets of Fig. 7. A new peak at 2θ = 28.89 as d = 1.55 Å appears from 200 K and its intensity linearly increases with cooling. The peak is a face centered orthorhombic (fco) - $NbH_{0.75}$ (220) and remains stably down to 50 K. At 100 K and 50 K, a weak peak associated with alpha (α)-Nb-H solid solution is vaguely seen at 2θ = 18.46 (d = 1.69 Å), on the low angle side of the Nb (200) plane. Similarly, all Nb-H phase peaks completely disappeared when the sample returned to 300 K.

**Surface depth profile of hydrogen and oxygen concentrate by SIMS (secondary ion mass spectroscopy)**

Secondary ion mass spectroscopy (SIMS) was carried out to locate the sources of hydrogens leading to the segregations of the nanometric scale Nb hydrides. Depth profiles of the hydrogen and oxygen contents near the surfaces of the hot spot of the non-degassed (Fig. 2) and the cut out of the H-enriched sample (Fig. 6) at room temperature are compared, as shown in Fig. 8. Hydrogen content increases to a maximum value immediately underneath the $Nb_2O_5$ layer in both samples, but the H-enriched sample exhibits 50% higher counts than the hot spot sample. While in both samples the oxygen counts peak at the surface and decay in direction of the Nb matrix, compared to the H-enriched sample, the oxygen counts in the hot spot cut-out decays more gradually. This suggest that some of the oxygen atoms from $Nb_2O_5$ on the surface of the hot-spot sample diffuse into the Nb matrix[36]. On the other hand, the formation of the $Nb_2O_5$ layer on the H-enriched sample depends heavily on the oxygen present in the conventional water lubricant-based mechanical polishing method used for the sample preparation[37]. This suggests that the presence of a large concentration of oxygen interstitial atoms in the Nb matrix for the non-degassed sample likely explains the relatively lower hydrogen content and weaker Nb-H precipitate signal observed.

**Discussion: Heterogenous segregation of nm-size $Nb_{1-x}H_x$ phase**

Hydrogen atoms behave as a source of hydride precipitate in the gas-metal (H-M) environment[38,39]. This is of upmost importance to superconducting radio frequency (SRF) technology using niobium[18,40,41] because the Nb hydrides can lead to thermal breakdown of superconductivity as a normal conductor[32,42]. In the case of H-enriched samples, micrometer-size of the Nb-H phase segregations are favorable, thereby, resulting in extended lattice distortions such as a scar on the surface along with destruction of the topmost dense $Nb_2O_5$ layer (in the supplementary Fig. 2). However, such deformation is not feasible in the practical SRF Nb cavities, even for the cases of "H-disease" or "HFQS" where RF performance markedly drops as described in the supplementary Fig. 1 and Fig. 2. Apparently, Nb-H precipitates in a nanometric scale play a more important role in such degradations[18]. If Nb-H precipitate has a length scale comparable to $\xi_{nb}$ (0) ≈40 nm, local degradation of superconductivity is induced by proximity effect, as a result, enhancement of cavity performance can be restrained.

Our first cryogenic-AFM scan presents precipitation of surface topographical features upon cooling, associated with the segregation of nanoscale Nb hydride precipitates on the hot spot of the non-degassed cavity. Nano-sized surface features appeared on the surface at 150 K, remained down to 10 K, and then completely vanished when warmed up to room temperature. The GI-XRD study suggests that bcc-NbH$_{0.8}$ phase segregation is clearly apparent in this sample across this same temperature range. The weak peak intensities, however, suggest that Nb-H phase segregations are restricted to the Nb surface layer. Similar to the cryo-AFM scan, the peaks of the Nb hydride on 2θ XRD patterns disappeared when warmed back to room temperature. This finding suggests that the nanometer-sized surface features likely arise from Nb-hydride precipitation.

Based on the findings with the hot spot of the non-degassed cavity (no heat treatment effect on the surface), "HFSQ" is obviously driven by substantial precipitations of the surface features that are associated with nano-size hydride segregations. However, the "HFQS" appeared on the EP'ed + 800°C/2hrs HT'ed cavity (supplementary Fig. 2) could be different because 800°C/2hrs heat treatment significantly reduced the feature formation, but with high density of smaller features, as shown in Fig. 3-5. GI-XRD study still shows a marked increase of the peak intensity of fco-Nb-H phase on the EP'ed + 800°C/2hrs HT'ed cavity from 200 K to 50 K, with a weak α-Nb(H) peak. As the H-related surface condition of the hop spot with post 800°C HT is supposed to be equivalent to the EP'ed + 800°C/2hrs HT'ed cavity, the total amount (or density) of nano-size Nb-hydride precipitations is the most decisive parameter for the extent of HFQS in the SRF Nb cavity performance.

The cut-outs from the 120°C baked, N-doped, and N-infused cavities show fewer and much smaller nano-size Nb hydride-like surface features, and these cavities do not show HFQS behavior. Thermal treatment at mild temperature (≤ 120°C) facilitates diffusion of oxygen (O) toward the Nb matrix[36]. Treatment with N (nitrogen) gas promotes interstitial invasion of N atoms into the cavity surface[43]. Since trapping of hydrogen atom is favorable with O (oxygen) and (N) nitrogen[40,44,45], it is not surprising to see the lowest density of surface features on the 120°C baked, N-doped, and N-infused cavity cut-outs.

By comparing the H-enriched environment with simulated Nb hydride peaks in cryogenic GI-XRD, we were able to distinguish the nano-sized Nb-hydride precipitates from the extensive irreversible surface deformation (in the supplementary Fig. 3). This shed light on the formation mechanism of nanometric Nb hydrides. At ambient temperature, hydrogen atoms are mobile in the Nb lattice. Redundant hydrogen atoms segregate at interstitial (tetragonal) sites or at defects such as voids[46], grain boundaries[47,48], or dislocations[49,50]. When the temperature is decreased below 300 K, hydrogen atoms diffuse to neighboring defects and Nb hydrides starts to precipitate locally. The finding that nanoscale Nb hydrides completely disappear when heated to 300 K suggests that the segregated Nb-H phases dissolve thoroughly during this process, and H atoms diffuse back to neighboring tetragonal sites. The two-state model[26,27] describes H concentration as being highest near the surface (<100 nm) rather than the bulk, and hydrogen atoms cannot diffuse through the $Nb_2O_5$ layer. Thus, if Nb-H precipitates form at the interface between the $Nb_2O_5$ oxide layer and Nb matrix and those hydrides protrude through the oxide layer, we can observe the nanoscale features with AFM scan. Figure 9 presents the schematic diagram of the heterogenous segregations of Nb hydride ($Nb_{1-x}H_x$) precipitates, proposed based on this scenario. Secondary ion mass spectroscopy (SIMS) shows a few nm thick $Nb_2O_5$ oxide layer present on the top layer of the H-enriched sample that exhibits both irreversible (micro-sized) and reversible (nano-sized) Nb-H precipitates. Observation of the nanoscale Nb hydrides could not be made by extended diffusion of hydrogen atoms via the thick oxide layer. Thus, it could be protrusions by Nb hydride segregation at the interface. SIMS depth profile (Fig. 8) presents that access hydrogens are mainly placed just underneath of the $Nb_2O_5$ layer. From the calculation of unit cell parameter based on the GI-XRD scan, slight hysteresis was found on lattice parameter after a cycle of cooling down and up. However, in the case of nanoscale hydride precipitates, the cooling cycle produced the reversible lattice structure changes.

**Conclusion**

From a systematic study by combining cryogenic atomic force microscopy (AFM) with high energy grazing incidence X-ray diffraction (GI-XRD), we were able to make the first direct observation on nanometric surface features associated with hydride precipitates on the SRF Nb cavities (3D resonators) upon cryogenic cooling. We found that such nanometer sized Nb hydrides are responsible for degradation of RF performance resulting from local breakdown of superconductivity by proximity effect. Importantly, our study suggests that the amount (density) of nano-Nb hydride precipitation is the most decisive parameter. With the surface depth profile reconstructed by secondary ion mass spectroscopy (SIMS), it is postulated that nanoscale surface pop-ups upon cryogenic cooling are developed by hydrides segregated at the interface between the $Nb_2O_5$ oxide layer and the Nb matrix. From the fact that the surface bumps completely disappeared without any detectable topographical features when warming back to the room temperature, it is feasible that the precipitated Nb hydrides completely dissolved at 300 K, and H atoms diffused back to interstitial sites or defects. Hydrogens can lead to distortion of the lattice structure through re-arrangement during a cycle of cooling and warming, thereby disordering of superconductivity can be also developed by hydrogen displacement. This study shows that the formation morphology of nanometric Nb hydrides varies with the engineering processes applied, suggesting that the extent of proximity degradation of RF superconductivity can be determined by post surface or thermal treatments to minimize hydrogen intakes.

**Method**

**Samples: the cut-out coupons from the tested SRF Nb cavities**

10 mm diameter of disk shape coupons were extracted from the representative superconducting radio frequency (SRF) cavities (3D resonators), as described in Table 1. Extraction was carried out by very slow drill-machining without water-based lubricant, in order to minimize hydrogen intrusion and structural deformation by cutting heat load. Cut-out coupons are slightly curved (concaved) shape due to the intrinsic circular pillbox shape of the cavity. The term of "hot spot" represents a local area of the cavity surface where breakdown of superconductivity occurs due to thermal fluctuation during RF operation[29]. No. 2 800°C HT'ed sample is from hot spot areas of the non-degassed cavity that was not thermally treated after a standard ~40 μm EP reset process. After extraction, the sample was thermally treated at 800°C for 2 hours and then ~ 20 min BCP (buffered chemical polishing) processed to remove surface damaged layers (~ 20 μm). No. 6 H (hydrogen)-enriched sample was prepared from a hot spot of a post 975°C HT + N-doped (at 800°C) Nb cavity with applying conventional cross section mechanical polishing, finished with vibrometry polishing in $H_2O$ based colloidal silica solution. This H-enriched sample preparation is identical to the recipe applied in the initial *in-situ* cryogenic study[24,25] with laser confocal microscopy. Nb hydrides precipitation on this H-enriched sample left behind permanent physical deformation like scars, as shown in SEM image and EBSD data in the supplementary Figure 3.

**Cryogenic Atomic Force Microscope (AFM)**

The cryogenic AFM applied is a commercial instrument of attocube system AG, which is incorporated with QD (quantum design) 9 Tesla (T) physical property measurement system (PPMS) that enables to cool a sample down to 1.8 K. To maximize topographical resolution, contact mode AFM was implemented with minimal surface loading force using a Nanosensors AFM tip of 0.022-0.77 N/m stiffness. 20 K/min cooling rate is applied, but ~3 hours staying duration at the target temperature was necessitated due to high level noise from thermal fluctuation of a cantilever piezo. The total scanning area reduce with decreasing of the system temperature because the movement of the piezo blocks decrease with temperature.

**High energy X-ray diffraction at grazing incidence**

High energy X-ray diffraction was performed at the sector 33-BM-C beamline of the Advanced Photon Source (APS) of ANL (Argon National Laboratory) facility, which delivers high flux and monochromatic beam using a double Si (111) monochromator. The high-resolution is achieved through the Huber 6-cicle diffractometer and Pilatus 100 K detector, with 900 μm x 500 μm as focused point. All the samples were cooled using a liquid helium cold finger (300 K to 30 K) under ultra-high vacuum. The selected energy used for obtaining the patterns was 16 eV ($\lambda$= 0.775 Å) and incident angle of 1º. The depth penetration by the X-rays (E= 16 keV) was calculated from the Parratt equation[51] resulting in ~0.75 μm. Peaks of 2θ patterns were indexed by applying DicVol integrated in FullProof software[52] with ICSD database to corroborate the phase identification[53]. The Nb-hydride $NbH_{0.8}$, ICSD number 638368, is found as a bcc (body centered cubic)-structure with a lattice parameter (å) of 0.342 nm as a $Im\bar{3}m$ space group. Orthorhombic (orth)-$NbH_{0.89}$ is the face centered orthorhombic structure, ICSD number 150604 and lattice parameters of a= 0.484, b= 0.49 nm and c= 0.345 nm with *Pnnn* as space group.

**Acknowledgement**

This manuscript has been authored by Fermi Research Alliance, LLC under Contract No. DE-AC02-07CH11359 with the U.S. Department of Energy, Office of Science, Office of High Energy Physics. This research used resources of the Advanced Photon Source; a U.S. Department of Energy (DOE) Office of Science User Facility operated for the DOE Office of Science by Argonne National Laboratory under Contract No. DE-AC02-06CH11357. Special thanks to Dr. Balachandran for thoughtful review.

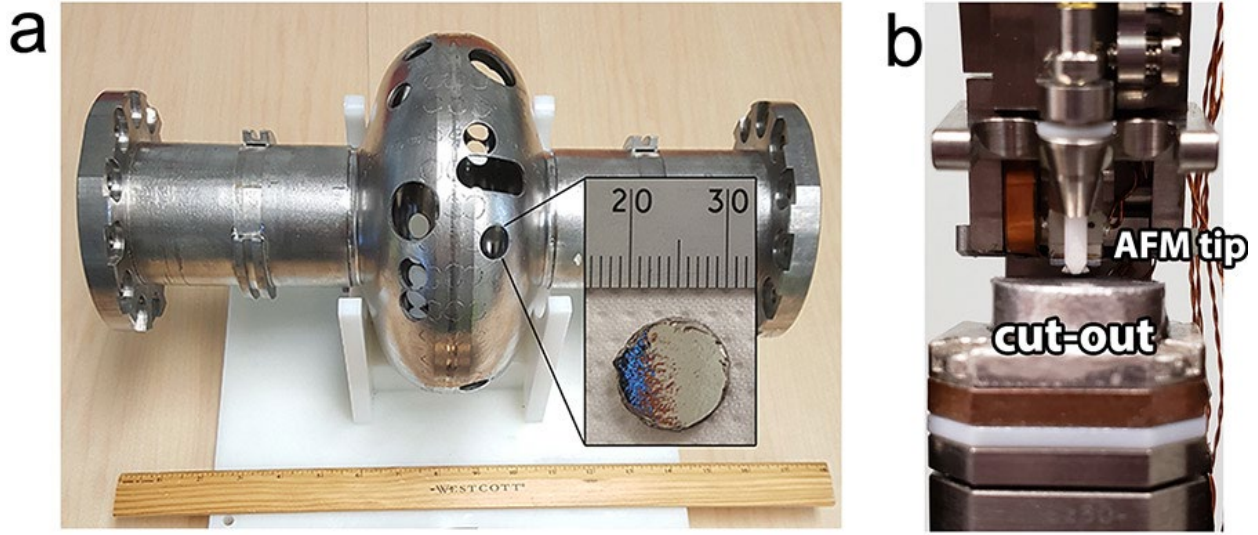

**Fig. 1 Single-cell Superconducting Radio Frequency (SRF) niobium cavity and Cryogenic AFM system. (a)** A photo of a single-cell SRF Nb cavity with an enlarged image of a cut-out coupon (10 mm of diameter and 2.5 mm of thickness). Several machined areas are seen on the outside of the cavity, where thermal heating occurred during a RF test, being described as breakdown of surface superconductivity. The ruler is 28 inch long. **(b)** A photo of atomic force microscope (AFM) with a cut-out sample on the sample stage of the piezo driven X-Y-Z stage block.

Table 1: Descriptions of the characteristics of the SRF Nb cavities and its cut-out coupons applied in this study

| No. | Cut-out Sample ID | Applied Engineering Features for Cavity and Cut-out coupon | RF Performance Characteristics |
|---|---|---|---|
| 1 | Hot spot | Non-degassed EP cavity: no thermal treatment and only extra 10 μm EP applied after a standard ~40 μm EP reset process | Q drop near 30 MV/m showing HFQS[31] |
| 2 | 800°C HT'ed | A hot spot cut-out coupon from the non-degassed EP cavity (No. 1). This coupon was thermally treated at 800°C/2hrs + 20 μm BCP after extraction | Surface condition regarding H-content is analogous to the EP + 800°C HT'ed cavity |
| 3 | 120°C Baked | EP + 120°C/48hrs baked cavity after 800°C/2hrs heat treatment | Up to 40 MV/m without HFQS |
| 4 | N-Doped | N (nitrogen)-doped cavity: 25mTorr of $N_2$ gas at 800°C for 2 min + 20 μm EP after 800°C/2hrs heat treatment. | Anti Q slope behavior up to the middle accelerating field regime (25-30 Mv/m)[12] |
| 5 | N-Infused | N (nitrogen)-infused cavity: 25mTorr of $N_2$ gas at 120°C for 48hrs after 800°C/2hrs heat treatment | Up to 45 MV/m without HFQS[14] |
| 6 | H-loaded (enriched) | Cross-sectionally mechanical polished hot spot taken from a post 975°C/3hrs heat treated N (nitrogen)-doped cavity with 25mTorr of $N_2$ gas at 800°C for 2 min | The 975°C HT'ed N-doped cavity showed meaningful flux expulsion behavior[35] |
| 7 | EP + 800°C HT'ed | 800°C HT'ed EP cavity: 800°C/3hrs heat treatment + 40 μm EP | Q drop near 30 MV/m showing HFQS[10] |

\* EP and BCP stands for electropolishing and buffer chemical polishing, respectively, which are representative surface chemical treatments for SRF Nb cavity fabrication in order to remove damaged surface structures. EP is superior to BCP in surface roughness[5].

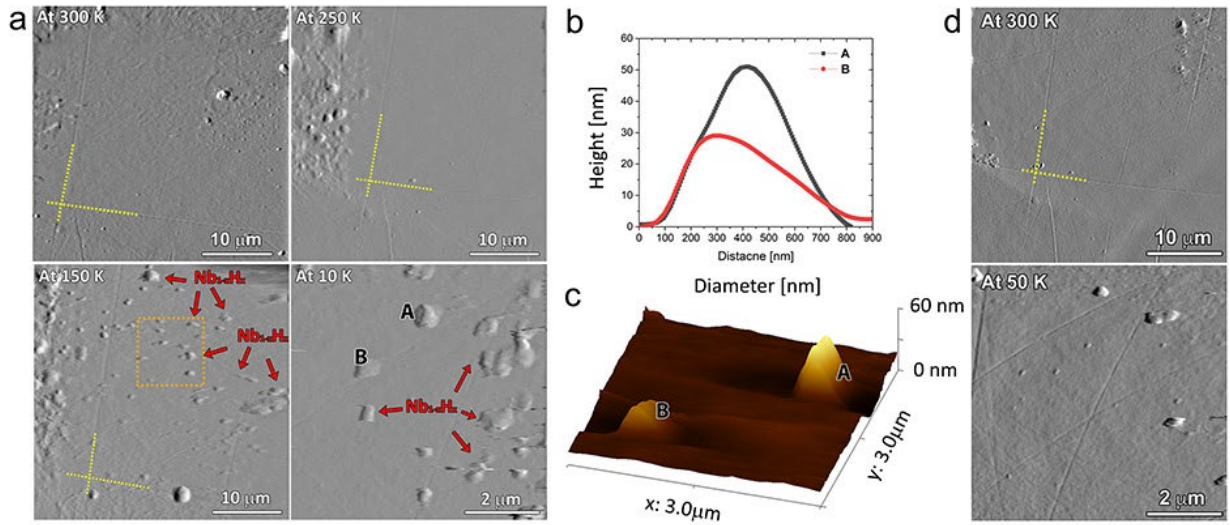

**Fig. 2 Cryogenic AFM images on the hot spot cut-out coupon from the non-degassed EP'ed (electropolished) cavity. a** A series of the AFM intensity images acquired during cooling from 300 K to 10 K, surface scratches marked with the yellow dotted lines are to give a guidance for locating the scan area. Distinct surface features that are suspected to be due to hydride precipitations appeared at the 150 K, which are indicated by the orange dotted square. This square area is enlarged at the 10 K. **b** Height profile across two bumps (A & B) marked on the intensity AFM image at 10 K. **c** 3D topography of the two bumps, **d** The intensity images after warmed back to 50 K and 300 K, respectively.

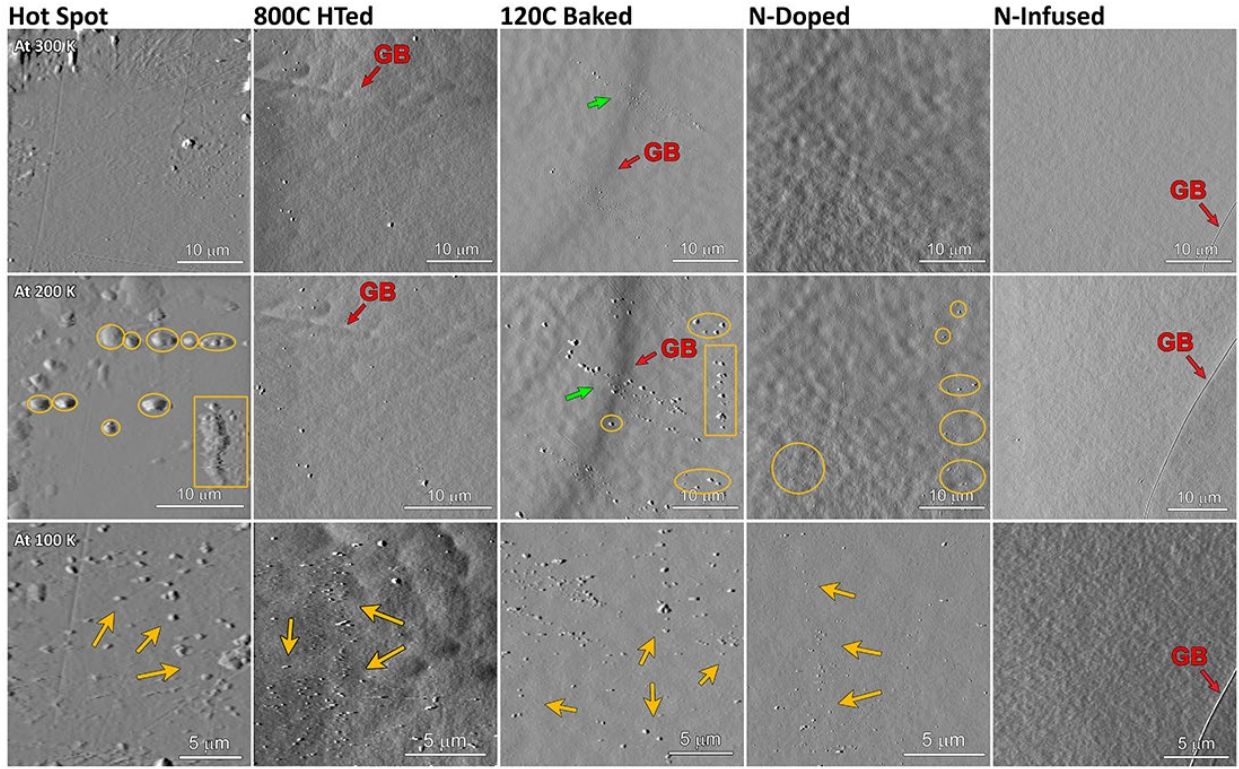

**Fig. 3. A series of laser-intensity AFM images of the cut-out coupons from the differently treated SRF Nb cavities at cryogenic temperature from 300 K to 100 K.** The cut-out coupons applied in this study are from the non-degassed (no heat treatment after a standard reset), the 120°C/48hrs baked (after 800°C/2hrs HT), N (nitrogen)-doped (800°C/2min with 25mTorr $N_2$ gas after 800°C/2hrs HT), and N (nitrogen)-infused (120°C/48hrs with 25mTorr $N_2$ gas after 800°C/2hrs HT) SRF Nb cavities, as described in Table 1. The hot spot is the same as the Fig. 2, and the 800°C HT'ed sample is one of hot spot areas of the non-degassed EP'ed cavity, post-annealed at 800°C for 2hrs and 20 μm surface chemical removed by BCP. Orange squares and circles indicate distinct surface features that appeared during cryogenic cooling. Grain boundaries (GBs) are marked with the red arrows.

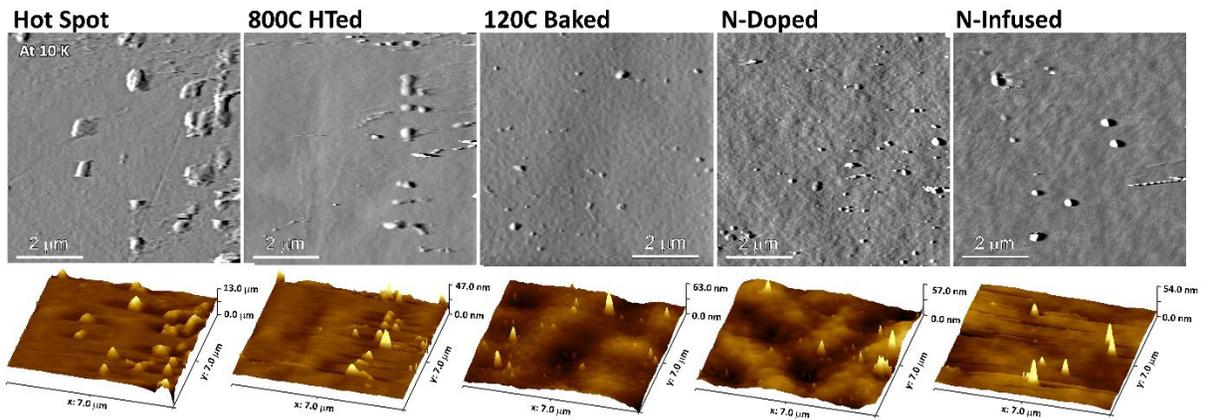

**Fig. 4. Comparison of surface topographical features on a scan of 7 μm by 7 μm area at 10 K.** AFM laser intensity images (top) and 3 D topography plots (bottom), acquired from the cut-outs of the five different SRF Nb cavities, same as Fig. 3.

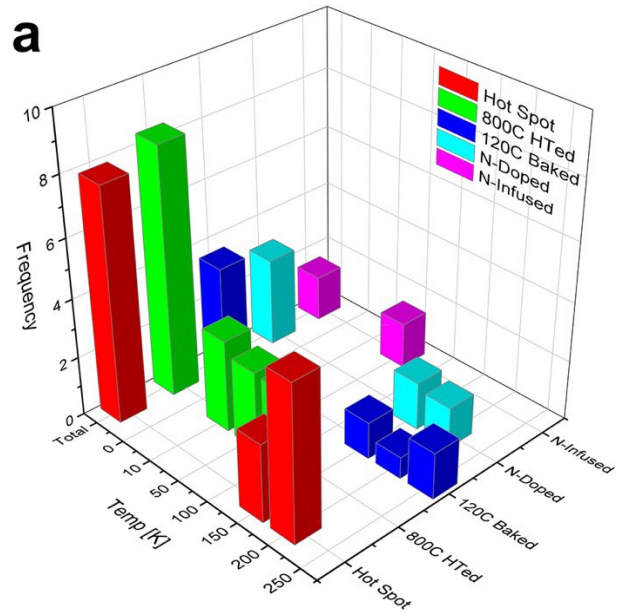

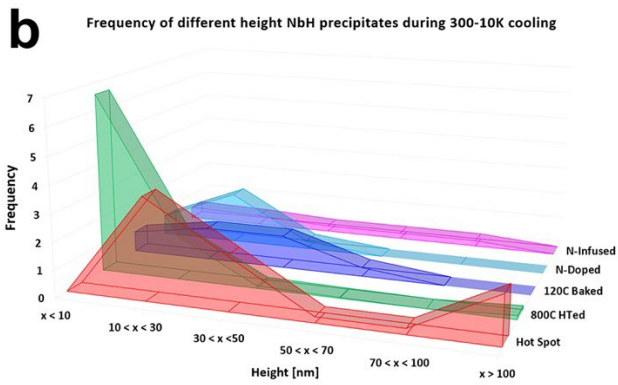

**Fig. 5.** The averaged frequency (density) of the surface pop-up features on the cut outs of the representative cavities over a unit of 37 by 37 μm² area during cryogenic cooling from 300 K to 10 K, **a** The frequency along with cooling temperature and **b** The frequency along with the height of the surface protrusions.

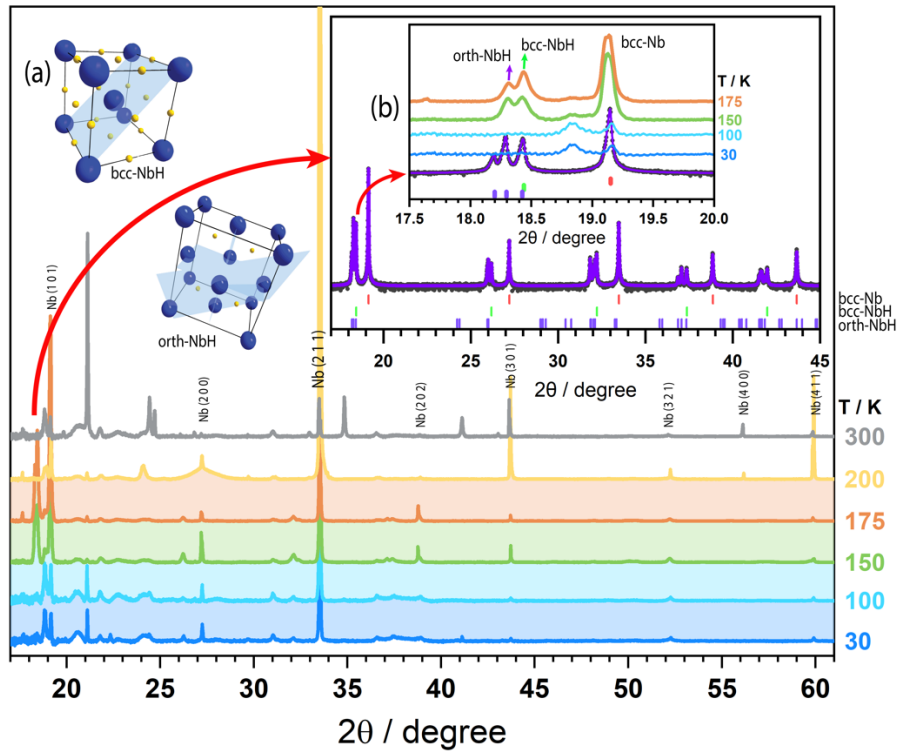

**Fig. 6. High-resolution XRD (X-ray diffraction) reflections from 300 K to 30 K.** 2θ X-ray pattern was acquired at 1° of the incident angles on the cross-sectionally mechanical polished (hydrogen (H)-enriched) cut out coupon, extracted from a hot spot area of a N-doped (at 800°C) cavity followed by post 975°C/3hrs heat treatment. Insets: **(a)** schematic unit cell structures of bcc-NbH$_{0.8}$ and orth-NbH$_{0.89}$ phases and **(b) Top:** the orange (175 K), green (150 K), light blue (100 K), and blue (30 K) lines: are experimental 2θ pattens at lower angle size of Nb (101) peak. **Bottom:** the violet line is the simulated 2θ pattern of the bcc-Nb (red bar), and bcc (green bar) and orthorhombic (violet bar) Nb-hydrides.

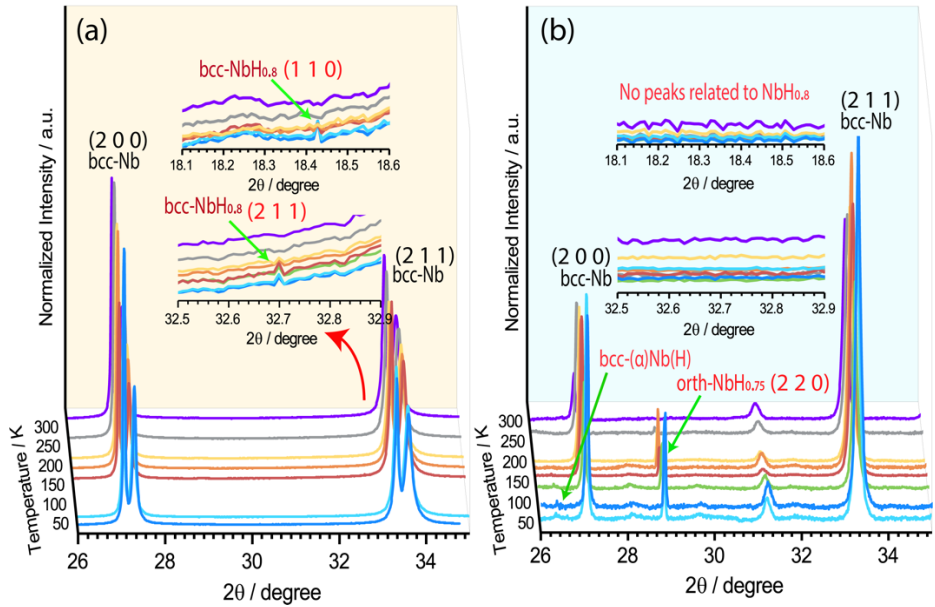

**Fig. 7. High-resolution XRD (X-ray diffraction) reflections at 1° of grazing incidence from 300 K to 50 K**. The 2θ X-ray patterns were recorded from (a) the hot spot of the non-degassed cavity, main bcc-Nb (110) and (220) peaks, inset: the peaks ascribed to bcc-NbH$_{0.8}$ (110) and (211) planes, (b) a cut out of the EP + 800°C heat treated cavity, bcc-Nb (110) main peak with α-Nb(H) and fco-NbH$_{0.75}$ (220) plane marked, inset: the same 2θ regime of the (a) inset is added for direct comparison. The color codes of the measurement temperature in the insets are the same as the main figure.

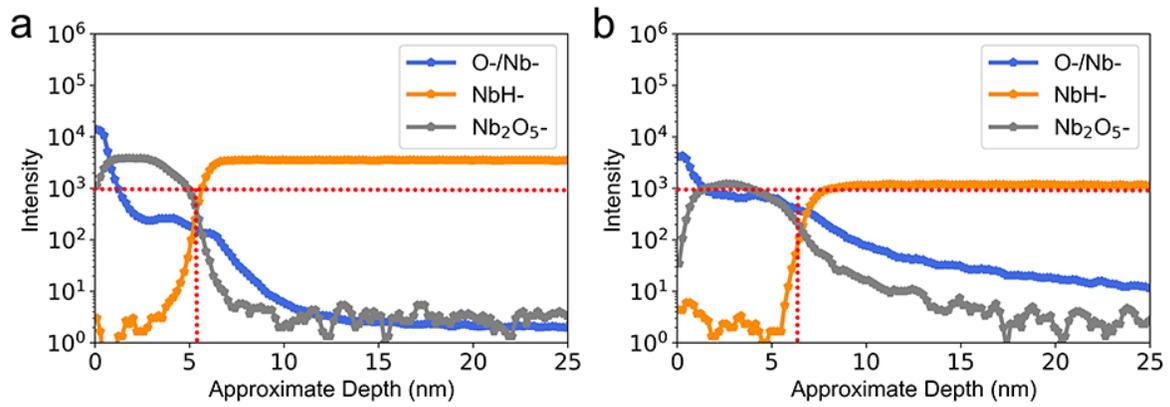

**Fig 8. Depth profiles of oxygens, hydrogen, and $Nb_2O_5$ by secondary ion mass spectroscopy (SIMS) on (a)** the H-enriched sample and **(b)** the cut out from the non-degassed cavity at 300 K. The red dotted line is provided for comparison.

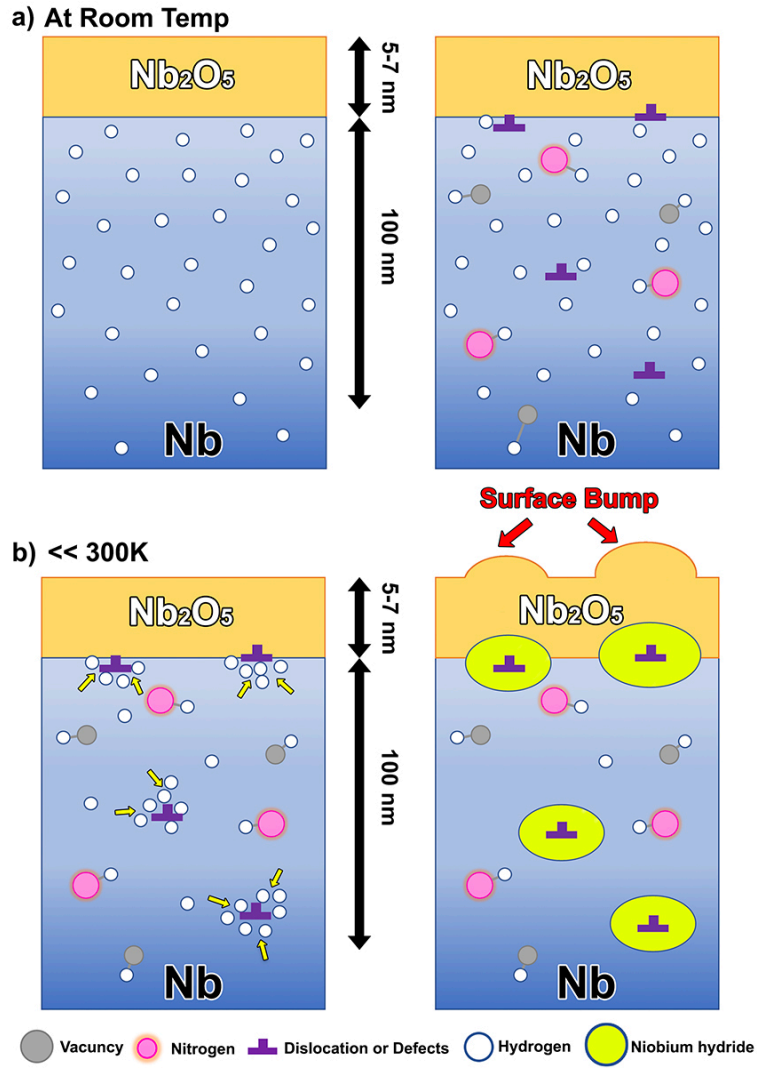

**Fig. 9.** Schematic diagram of the heterogeneous segregation of niobium hydride ($Nb_{1-x}H_x$) precipitates at the interface between the $Nb_2O_5$ surface oxide layer and the Nb matrix.

**Supplementary Data**

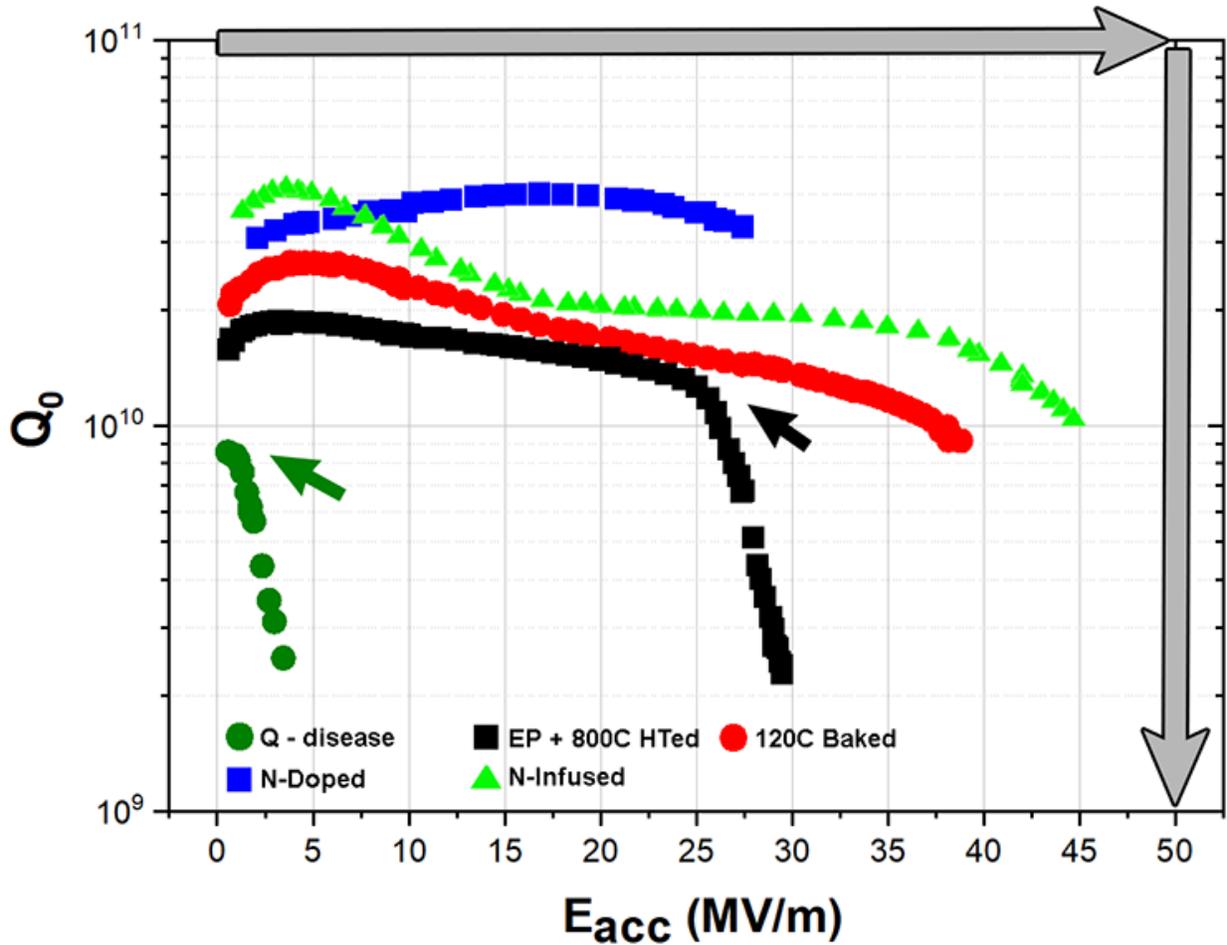

**Figure 1. Quality factor ($Q_0$) versus accelerating field ($E_{acc}$) of the representative of the superconducting radio frequency (SRF) niobium (Nb) cavities applied in this study**: 1) Q-disease, 2) 800°C HT'ed + EP (electropolishing), 3) EP (electropolishing) + 120°C baked, 4) N (nitrogen)-doped, and 5) N (nitrogen)-infused cavities. The applied engineering fabrication procedures for the cavities are described in Table 1. The green arrow indicates a marked Q drop due to hydrogen disease[31] and the black arrow points out the onset of Q degradation at high field RF regime, called high field Q slope (HFQS). The gray arrows describe the maximum theoretically expected performance of SRF Nb cavity up to the critical superheating field, $H_{sh}(0) \approx 250$ mT[5].

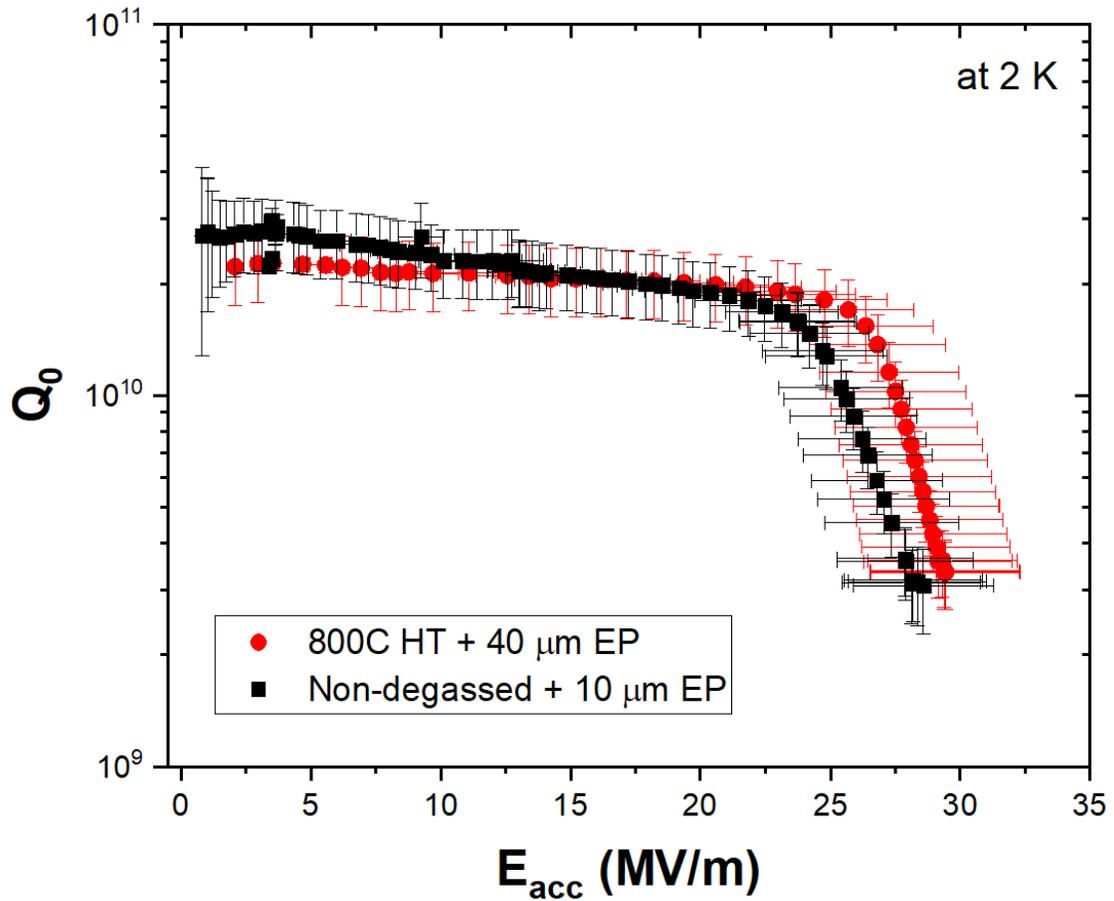

**Figure 2. Quality factor ($Q_0$) versus accelerating field ($E_{acc}$) of the superconducting radio frequency (SRF) niobium (Nb) cavities applied in this study**: black square) non-degassed EP cavity; no heat treatment applied and only 10 μm EP was performed, after the standard ~ 40 mm EP cavity reset process, and red circle) 800°C HT'ed EP cavity; 800°C/3hrs heat treatment and 40 μm EP applied.

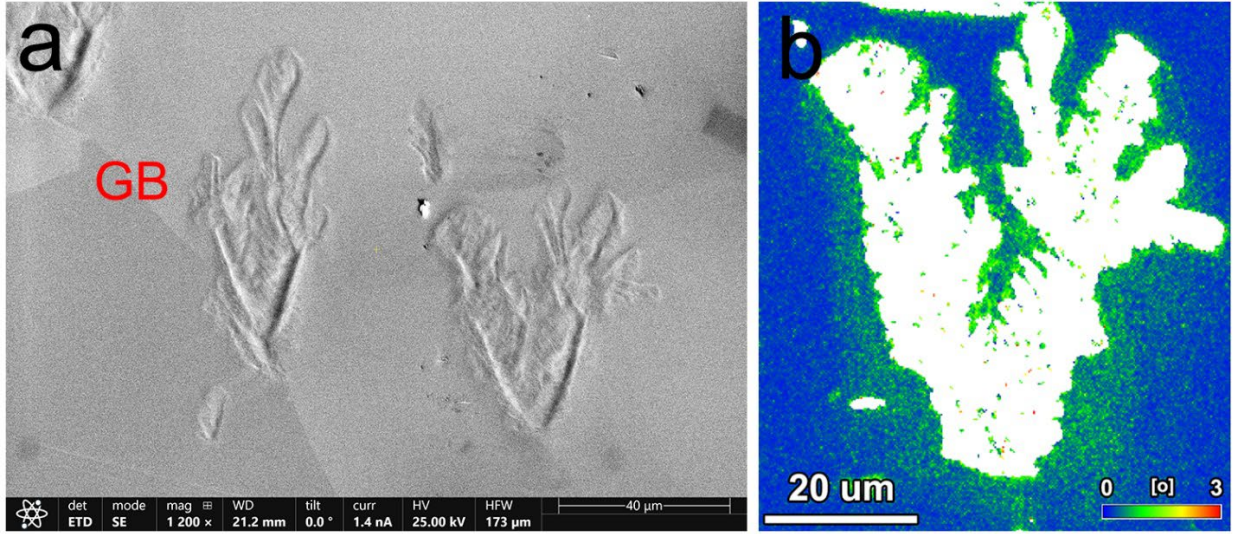

**Figure 3. (a) Back scattered electron (BSE) image and (b) Kernel misorientation map** of electron back scattered diffraction (EBSD) on the H-enriched sample after cryogenic XRD study at grazing incidence, which was extracted from the post 975°C/3hrs + N-doped (with 25 mTorr of nitrogen ($N_2$)-gas at 800°C/2min) SRF Nb cavity. Inset shows the misorientation scale (0-3 degree) bar.